\documentclass[12pt]{article}

\usepackage{epsfig}
\usepackage{comment}
\usepackage{cite}
\usepackage{amsmath, amssymb, amsfonts}
\usepackage{xcolor}
\usepackage{latexsym}
\usepackage{graphicx}
\usepackage{soul}
\usepackage[normalem]{ulem}
\usepackage[colorlinks,bookmarks]{hyperref}
\hypersetup{pdfpagemode=UseNone, pdfstartview=FitH, linkcolor=blue,
            citecolor=red, urlcolor=blue}

\def\be{\begin{equation}}
\def\ee{\end{equation}}
\def\ba{\begin{eqnarray}}
\def\ea{\end{eqnarray}}

\def\bdm{\begin{displaymath}}
\def\edm{\end{displaymath}}

\def\bq{\begin{quote}}
\def\eq{\end{quote}}

 at 10truept

\newcommand{\rmd}{\mathrm{d}}

\newcommand{\avg}[1]{\langle #1 \rangle}

\renewcommand{\[}{\left[}

\newcommand{\Om}{\Omega}

\newcommand{\mm}{\textrm{mm}}
\newcommand{\m}{\textrm{m}}
\newcommand{\km}{\textrm{km}}
\newcommand{\s}{\textrm{s}}
\newcommand{\eV}{\textrm{eV}}
\newcommand{\GeV}{\textrm{GeV}}
\newcommand{\nHz}{\textrm{nHz}}
\newcommand{\Hz}{\textrm{Hz}}
\newcommand{\kHz}{\textrm{kHz}}
\newcommand{\MHz}{\textrm{MHz}}

\newcommand{\Mpc}{\textrm{Mpc}}

\newcommand{\Mpl}{M_{\mathrm{Pl}}}

\newcommand{\fnl}{f_{\mathrm{NL}}}

\newcommand{\bea}{\begin{eqnarray}}
\newcommand{\eea}{\end{eqnarray}}

\newcommand{\bi}{\begin{itemize}}
\newcommand{\ei}{\end{itemize}}

\newcommand{\beq}{\begin{equation}}
\newcommand{\eeq}{\end{equation}}
\newcommand{\beqa}{\begin{eqnarray}}
\newcommand{\eeqa}{\end{eqnarray}}

\def\ltap{\ \raise.3ex\hbox{$<$\kern-.75em\lower1ex\hbox{$\sim$}}\ }
\def\gtap{\ \raise.3ex\hbox{$>$\kern-.75em\lower1ex\hbox{$\sim$}}\ }
\def\gl{\ \raise.5ex\hbox{$>$}\kern-.8em\lower.5ex\hbox{$<$}\ }
\def\roughly#1{\raise.3ex\hbox{$#1$\kern-.75em\lower1ex\hbox{$\sim$}}}

\graphicspath{{plots}}

\begin{document}

\thispagestyle{empty}
\begin{flushright}
January 2026\\
DESY 25-114  \\
\end{flushright}
\vspace*{.2cm}
\begin{center}

{\Large \bf Very-High-Frequency Gravitational Waves\\[0.3cm]  from Multi-Monodromy Inflation}

\vspace*{.7cm} {\large Guido D'Amico$^{a, b,  }$\footnote{\tt
guido.damico@unipr.it}, Andrew A. Geraci$^{c, }$\footnote{\tt
andrew.geraci@northwestern.edu}, Nemanja Kaloper$^{d, }$\footnote{\tt
kaloper@physics.ucdavis.edu} and \\}
\vskip.3cm
{\large 
Alexander Westphal$^{e,}$\footnote{\tt alexander.westphal@desy.de}
}\\
\vspace{.5cm} {\em $^a$Department of Mathematical, Physical and Computer Sciences}
\vspace{.05cm}{\em University of Parma, 43124 Parma, Italy}\\
\vspace{.2cm}
{\em $^b$INFN Gruppo Collegato di Parma, 43124 Parma, Italy}\\
\vspace{.2cm}
{\em $^c$Center for Fundamental Physics, Department of Physics and Astronomy \\ Northwestern University, Evanston, IL 60208, USA}\\
\vspace{.2cm}
{\em $^d$QMAP, Department of Physics and Astronomy, University of
California}\\
\vspace{.05cm}{\em Davis, CA 95616, USA}\\
\vspace{.2cm} $^e${\em Deutsches Elektronen-Synchrotron DESY, Notkestr. 85,}\\
\vspace{.05cm}{\em 22607 Hamburg, Germany}\\

\vspace{.7cm} ABSTRACT
\end{center}
We show that in multi-stage axion monodromy inflation an 
interruption near the end of the penultimate stage can lead to a spike in the 
gravitational wave background.
These gravitational waves are in the frequency range and with an 
amplitude accessible to proposed terrestrial detectors such as the 
Einstein Telescope, Cosmic Explorer, and future Levitated Sensor Detector experiments.

\vfill \setcounter{page}{0} \setcounter{footnote}{0}

\vspace{1cm}
\newpage

\section{Introduction}
\label{sec:intro}
\vskip.3cm
    
Our Universe is big, old, and very smooth at the largest scales.
The inflationary paradigm \cite{Guth:1980zm,Linde:1981mu,Albrecht:1982wi} is the 
leading candidate for explaining these facts in the framework of local causal theory of matter coupled to gravity.
The idea was that at some early time and high energy, the universe underwent a stage of 
accelerated expansion, which blew it up quickly in a short period of time. To fit the observations 
of the late universe in the region which evolved from inside a small causally connected patch, 
the expansion should last about $\sim 60$ or so efolds.
Typically, the models proposed to be behind such dynamics strive to accomplish 
it in one go, positing that there was a single such stage where all the required 
rapid expansion happened, after which the agent that drove the cosmic acceleration 
decayed into normal matter, allowing the expansion to slow down and smoothly 
connect to a radiation and matter dominated universe at late times.
This decay process, or ``reheating", occurs at the end of inflation, 
and it involves complicated dynamics, but their signatures, if any, are at 
scales which today are typically contaminated by dynamics that came much later than inflation.

Realizing inflation in field theory is not so simple, however. 
Explaining the dynamics which yields almost-vacuum energy
domination for a period of $\sim 60$ efolds and then disappears in 
the firework of particle production  ending inflation
requires nontrivial field dynamics, often needing scalar fields 
varying over (super-)Planckian ranges \cite{Lyth:1996im,Efstathiou:2005tq}. Constructing
meaningful models which can accommodate this is challenging in UV-complete frameworks, 
and at the very least it requires many
ingredients with initial conditions which ``conspire" 
to produce just the right environment (see, e.g. \cite{Kaloper:2011jz,eva}). Further, the direct
tests of inflation are few and far between: the geometry of inflation is locally de Sitter space, which by 
its enhanced symmetry serves as a cosmic ``amnesia tonic", erasing the initial conditions, and, typically, 
the field dynamics during inflation, as codified in the cosmic 
no-hair theorem \cite{Wald:1983ky,Kaloper:2018zgi}. Finally, many models are 
under pressure from the few observables which they produce, including the scalar power spectrum $\delta_{\tt S}$,
the scalar spectral index $n_{s}$ and the bounds on the primordial stochastic tensor background, parameterized
by the tensor to scalar spectrum ratio $r$. 

Yet, there is a simple variation of the conventional approach to inflation, 
which has been noted recently in
\cite{DAmico:2020euu}, 
that can meet some of these challenges head-on. As an added bonus, this approach 
can produce additional testable signatures, naturally avoiding the obstacle of the 
no-hair theorem. The idea is to have inflation 
occur in stages, driven by multiple inflaton sectors which, while similar to each other, are separated by
mass hierarchies (earlier discussions of various features of multi-stage inflation can be found in \cite{Kofman:1985aw,Starobinsky:1985ibc,Silk:1986vc,Mukhanov:1991rp,Polarski:1992dq,Pi:2019ihn,Damour:1997cb,Adams:1997de,Burgess:2005sb,Cicoli:2014bja}). 
Such frameworks can be realized within the context of braneworlds, with multiple copies
of inflatons arising from different branes, or from variants of the axiverse~\cite{Arvanitaki:2009fg}.
Some specific models were considered in \cite{Wald:1983ky,Kaloper:2018zgi}. 
Here we will not delve into the model building details, instead focusing on the novel observational
signatures they offer. In \cite{DAmico:2021vka,DAmico:2021fhz} 
it was already noted that in models where the inflatons are axions which couple 
to dark $U(1)$ gauge fields, near the end of the inflationary stage dominated by one such field, there can be an 
exponentially enhanced production of $U(1)$ waves, which in turn quickly 
convert to primordial gravitational waves with an
amplitude that greatly exceeds the usual stochastic gravity wave spectrum in a very narrow
range of wavelengths. Such gravitational waves occur at scales much shorter than the current horizon scale, and 
can be probed by an array of ongoing and planned surveys, such as NANOgrav, LISA, SKA and so on.

The stochastic gravitational wave backgrounds studied in \cite{DAmico:2021vka,DAmico:2021fhz} 
concern  gravitational waves with wavelengths of at least a 
few thousand kilometers, or with frequencies below $100 \, \Hz$. 
However, the interruptions during which they are
formed occur at least $10$ to $15$ efolds before the end of inflation. This immediately suggests that 
if there are interruptions even later into inflation, during those even shorter wavelength gravitational waves can
be induced. Indeed, a simple numerical estimate suggests that if the last 60 efolds of inflation began around
the GUT scale, when the Hubble parameter is about $10^{14} \, \GeV$, and the reheating is very efficient
producing the heat bath close to the GUT scale temperature, the scale factor of the 
subsequent decelerating universe would expand by a factor 
\be
\frac{a({\rm now})}{a({\rm reheating})} \sim \frac{T({\rm reheating})}{T({\rm now})} \sim 
\frac{10^{14} \, \GeV}{10^{-3} \, \eV} \sim 10^{26} \, ,
\label{efoldsa}
\ee
i.e. by another $\sim 60$ efolds\footnote{In this argument we are assuming that the interrupted universes are
radiation-dominated, following a very successful reheating after each interruption. This is for illustrative purposes  
only. If e.g. the interrupted epochs were matter-dominated the same argument would persist, the only change being 
some numerical modifications of the relationship between relic wavelengths, 
cosmological parameters and the efold where the interruption occurs.}. Thus a 
gravitational wave formed at  practically the end of this stage of inflation,
frozen in with a wavelength $\lambda \sim 1/H({\rm inflation}) \sim 10^{-14} \, \GeV^{-1} \sim 10^{-26} \, \mm$ 
would stretch by FRW decelerated expansion to about $\lambda({\rm now}) \sim \mm$ now.
Such short gravitational waves could in principle be easily contaminated by the production of 
gravitational waves in later universe, and it could be very difficult to separate 
them out from the backgrounds induced by local physics. 

\begin{figure}[ht]
    \centering
    \includegraphics[scale=0.5]{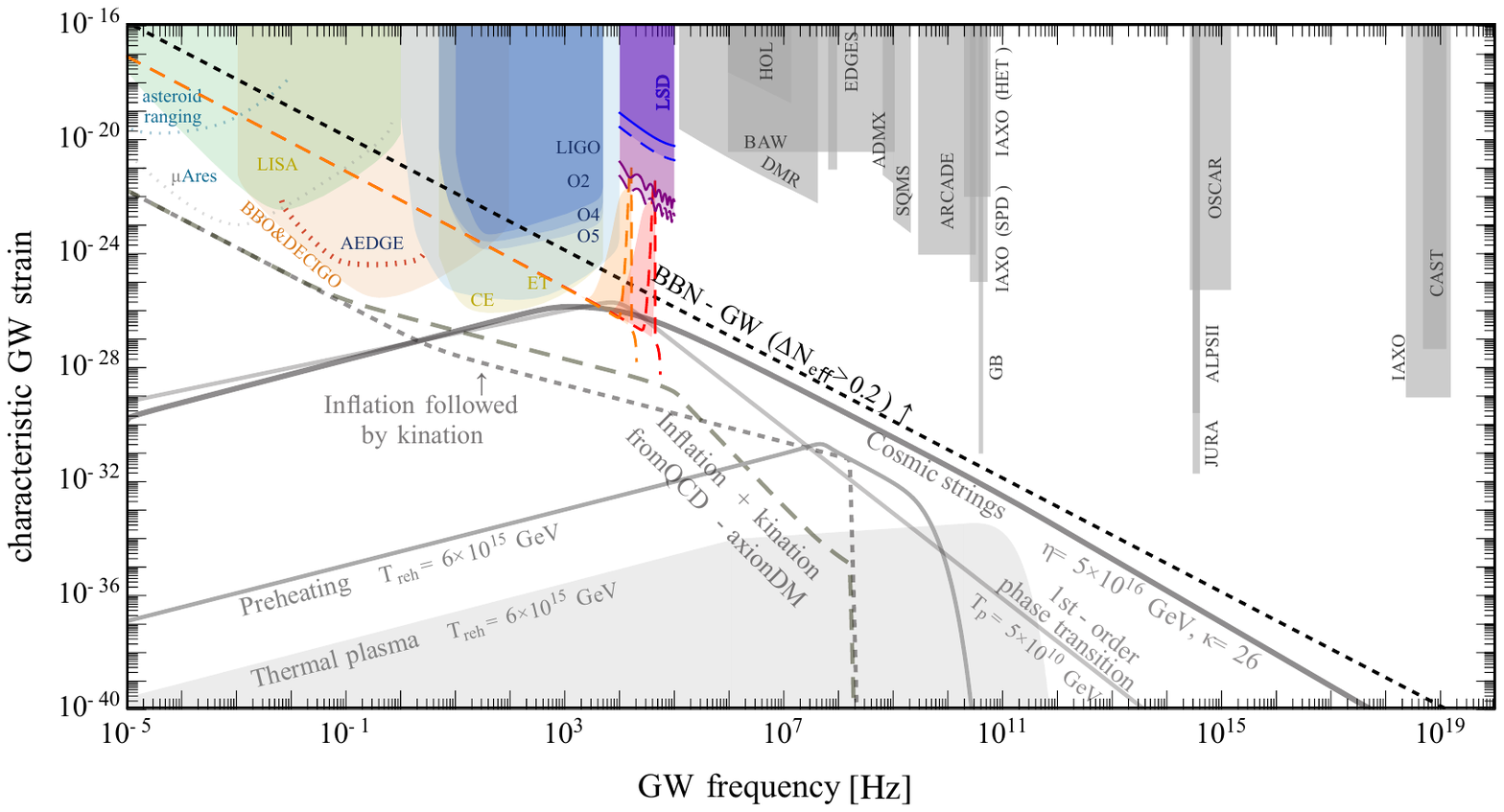}
    \caption{Gravitational wave strain as a function of frequency from different sources discussed in the literature, 
    such as cosmic strings, preheating after inflation, the hot thermal post-big bang plasma, etc. The bright red 
colored spikes near the center of the image are the signals derived in this paper. Clearly, if they were present in 
our universe they would dominate over other sources. Adapted from \cite{Servant:2023tua}.}
\label{fig:hGWvsN_sources}
\end{figure}

However, given the argument above, it is clear that it does not take much to shift the wavelength up and consider
gravitational waves which were produced somewhat 
earlier than the last efold of inflation. That would offer the possibility
to search for such gravitational waves in the range of wavelengths 
where the contaminations induced by local physics
happening in the late universe are small or absent. This is illustrated in Figure (\ref{fig:hGWvsN_sources}), 
which we adopted from \cite{Servant:2023tua}, and which clearly and efficiently compiles the findings of the works 
\cite{LIGOScientific:2016wof,Hild:2010id,LIGOScientific:2019vic,Punturo:2010zz,LIGOScientific:2014qfs,LISACosmologyWorkingGroup:2022jok,Aggarwal:2020olq,amaroseoane2017,Yagi:2011wg,AEDGE:2019nxb,KAGRA:2013rdx,Blas:2021mqw,Fedderke:2021kuy}, 
comparing the sensitivity of existing and planned experiments to the
anticipated gravitational wave signals from a variety of sources. We direct the reader to \cite{Servant:2023tua} 
for a detailed interpretative discussion of  Fig.\ref{fig:hGWvsN_sources}. Here we merely note that the 
signal which we consider may be powerful enough to exceed by orders of magnitude other sources, and
may therefore win hands down over them in terms of possibilities for detection. 

Indeed, once we restore the realistic cosmological parameters
which match inflationary prediction within the context of multi-monodromy models of inflation, we find that 
if an interruption occurs a few efolds (i.e. 6 to 10) 
before the end of the last 60 required to solve the flatness and homogeneity problems,
the chiral gravitational waves produced by the inflaton decaying into a dark $U(1)$ which converts to gravitational
waves have frequencies $\nu \sim 10^4 \, \Hz - 5 \times 10^5 \, \Hz$, which corresponds to wavelengths 
$\lambda \sim 30 \, \km - 600 \, \m$. 
This is in the domain that could be explored with terrestrial experiments 
such as Einstein Telescope or Cosmic Explorer ($\sim {\cal O}(10 \, \km)$) 
or extended versions of the Levitated Sensor Detector ($\sim {\cal O}(100 \, \m)$). 
It turns out that the amplitude of these gravitational waves is right around
the detection threshold of these instruments. This evades the bounds on the gravitational wave power
depicted in Fig.\ref{fig:hGWvsN_sources}, specifically both the limits coming from 
Big Bang Nucleosynthesis (BBN) and from the consideration of cosmic string sources, 
because our signal has a very narrow frequency range, and therefore
the total power of these gravitational waves, indicated by the area below the red spikes in 
Fig.\ref{fig:hGWvsN_sources}, is easily much smaller than the area below the BBN limiting curve and
the cosmic strings curve even when the red spike is as many as $\sim {\cal O}(10)$ orders of magnitude larger. 

This remarkable happenstance therefore offers a chance to search for signatures of 
inflationary cosmology, which while model dependent, follow from arguably realistic models of 
inflation from the point of view of UV physics. Combined with more interruptions, 
and corresponding gravitational waves
sourced with larger wavelengths, this offers a remarkable opportunity for searching for new physics from 
inflation with gravitational wave detectors. In our view this exciting possibility deserves serious and detailed 
exploration.

\section{Multi-monodromy inflation and the ``characteristic" spectrum of GW signals}

Axions are attractive candidates for driving cosmological slow-roll inflation, both from the 
bottom-up EFT point as well as within string theory. The reason is the discrete shift symmetry 
and the dual formulation gauge symmetry, 
which good axions enjoy, which in turn constrains corrections to the axion scalar potential to 
arise in powers of those operators involved in generating the leading-order axion potential 
in the first place. As this removes the appearance of arbitrary just $\Mpl$-suppressed 
monomial terms, the so-called eta problem is absent for axion inflation models. 

The scalar potential for axion inflation can arise as a manifestly periodic potential 
from non-perturbative quantum corrections breaking the continuous shift symmetry down 
to a discrete subgroup. Or the axion can acquire an infinite set of individually  non-periodic 
power-law scalar potentials from axion 
monodromy~\cite{McAllister:2008hb,Kaloper:2008fb,Dong:2010in,Kaloper:2011jz}. 
This infinite set of branches with their minima separated by $2\pi f$ realizes the 
discrete shift symmetry as a property of the full branch system of scalar potentials, 
even though the individual branch breaks the shift symmetry completely. 
Here, $f$ denotes the axion decay constant of the axion arising from its kinetic term, 
which in turn sets the periodicity scale $2\pi f$ of the non-perturbative quantum corrections.

We can get this core mechanism of axion monodromy both for the KK zero 
mode axions arising from higher $p$-form gauge fields of string theory compactifications 
in the presence of either branes or higher-dimensional quantized $q$-form field strengths, 
and from the unique 4D EFT description of a massive axion arising from mixing a 
pseudo-scalar with a non-dynamical but vacuum energy changing 4-form field 
strength. As the protection\footnote{The gauge symmetries do not completely tame all
UV dangers since the gauge field mass might be UV sensitive, as in e.g. hierarchy problem (we ignore the
cosmological constant problem). We assume that some mechanism in the UV completion 
of the theory fixes it to be sufficiently small, and then the gauge symmetries maintain
control of other operators \cite{Kaloper:2016fbr}.}
of the branch scalar potentials from most quantum corrections is much 
more transparent from the underlying gauge symmetries governing the 
4D effective massive 3-form gauge theory~\cite{Kaloper:2008fb,Kaloper:2011jz} (see also 
\cite{Dvali:2005an} for important precursor work), we will largely stick with this 
4D effective description
\begin{equation}
\frac{{\cal L}}{\sqrt{-g}}=-F_{\mu\nu\rho\sigma}^2
-(\partial \phi)^2 - m \phi \epsilon_{\mu\nu\rho\sigma} F^{\mu\nu\rho\sigma}\quad.
\end{equation}
Carefully integrating out the 4-form field strength 
$F_{\mu\nu\rho\sigma}=\partial_{[\mu}A_{\nu\rho\sigma]}$ shows the axion $\phi$ 
acquiring an infinite set of branches of quadratic scalar potentials $\sim m^2\phi^2$.

A generic feature of axion monodromy inflation is the dynamics of flattening - the branch 
scalar potential tends to flatten out below the tree-level $\phi^2$-behavior when $\phi$ is 
displaced by more than a distance $\mu = {\cal O}(0.1 -1) \, \Mpl$ from the branch minimum. 
This flattening is already contained in the 4D effective 
flux monodromy description \cite{Kaloper:2008fb,Kaloper:2011jz} once 
higher gauge-invariant operators are included. We will assume scaling 
relationships between the Wilson coefficients of these operators such that the 
whole series resums into a flattened power-law form visible in the existing simple 
string models of axion monodromy inflation. Given this background story, 
the effective scalar potential for an axion from axion potential on each branch takes the form
\begin{equation}
V(\phi)= M^4 \left[\left(1+\frac{\phi^2}{\mu^2}\right)^{p/2}-1\right]\quad.
\end{equation}
Here the scale $M$ is ultimately determined by fitting the CMB 
temperature anisotropy amplitude. For the asymptotic power $p$ of the 
scalar potential $V\sim \phi^p\;,\; \phi\to\infty$ we take guidance from the 
known top-down string examples which suggest $0.1\lesssim p < 1$.
As long as the axion has canonically normalized 2-derivative kinetic term, 
this largely fixes the CMB observable predictions to $n_s > 0.975$ and $ 0.02 < r < 0.08$, 
which is too blue an $n_s$ given the current constraints from 
data~\cite{DAmico:2017cda,Dias:2018koa,DAmico:2021vka}.
Axion monodromy inflation is a large-field model of inflation, and 
we see this here for the range of $p$ as well, where 60 efolds of 
inflation correspond to ${\cal O}(5 - 10)\,\Mpl$ of field range traversed. 

Clearly, the setup with a single axion with a 2-derivative kinetic term 
and the type of flattening motivated from string theory examples is under pressure 
from the data. We alleviate this by widening our view: 
\begin{itemize}
\item The gauge symmetries governing the 4D EFT of axion monodromy allow 
for higher-derivative kinetic terms of the axion. Examples constructed from 
adding a few such terms perturbatively as motivated from bulk $\alpha'$-corrections 
in string theory, or by adding an infinite series of such terms motivated from DBI action of 
D-branes, show that these higher-derivative terms will lower $r$ by decreasing the speed 
of sound $c_s$ of the curvature perturbation at the price of increasing the amount of 
equilateral non-Gaussianity $\fnl$. From the existing CMB bounds on $\fnl$, 
we infer that higher-derivative axion kinetic terms can suppress the 
lower bound on $r$ to about $r\gtrsim 0.006$.

\item A look at the structure of the six or seven extra dimensions used 
for compactifying string theory to 4D shows us that many choices have 
a large number of non-trivial subspaces (`cycles') within the compact extra 
dimensions. Together with the existence of higher $p$-form gauge potentials in 
string theory this leads to typical string compactifications having many axions 
as KK zero modes of those gauge potentials -- a string `axiverse'. 
\end{itemize}

Hence, there is no reason to assume that the observable 60 efolds of inflation 
have to come from continuous slow-roll of a single axion. Instead, several axions 
with each of them having a potential from monodromy may share the 
work~\cite{DAmico:2020euu,DAmico:2021vka,DAmico:2021fhz}. 

If, as will be typical without special arrangements, the scalar potential 
for each axion will have a somewhat different overall scale, inflation will then proceed in several stages.
Assume a small number $N$ of axions are displaced from their respective minima 
by several $\Mpl$. The heaviest axion will initial slow-roll, with the lighter axions being 
in very deep slow-roll essentially frozen in their smaller potentials. Once the heaviest 
axion rolls off into its minimum, it will break slow-roll for a short period of 2-3 efolds of 
red-shifting matter domination until the next-lighter axion's potential energy starts driving 
slow-roll inflation again once beginning to dominate. The story repeats with each 
subsequent lighter and displaced axion, leading to `roller-coaster' monodromy inflation. 

\begin{figure}[ht]
    \label{fig:Vinf}
    \centering
    \includegraphics[scale=0.4]{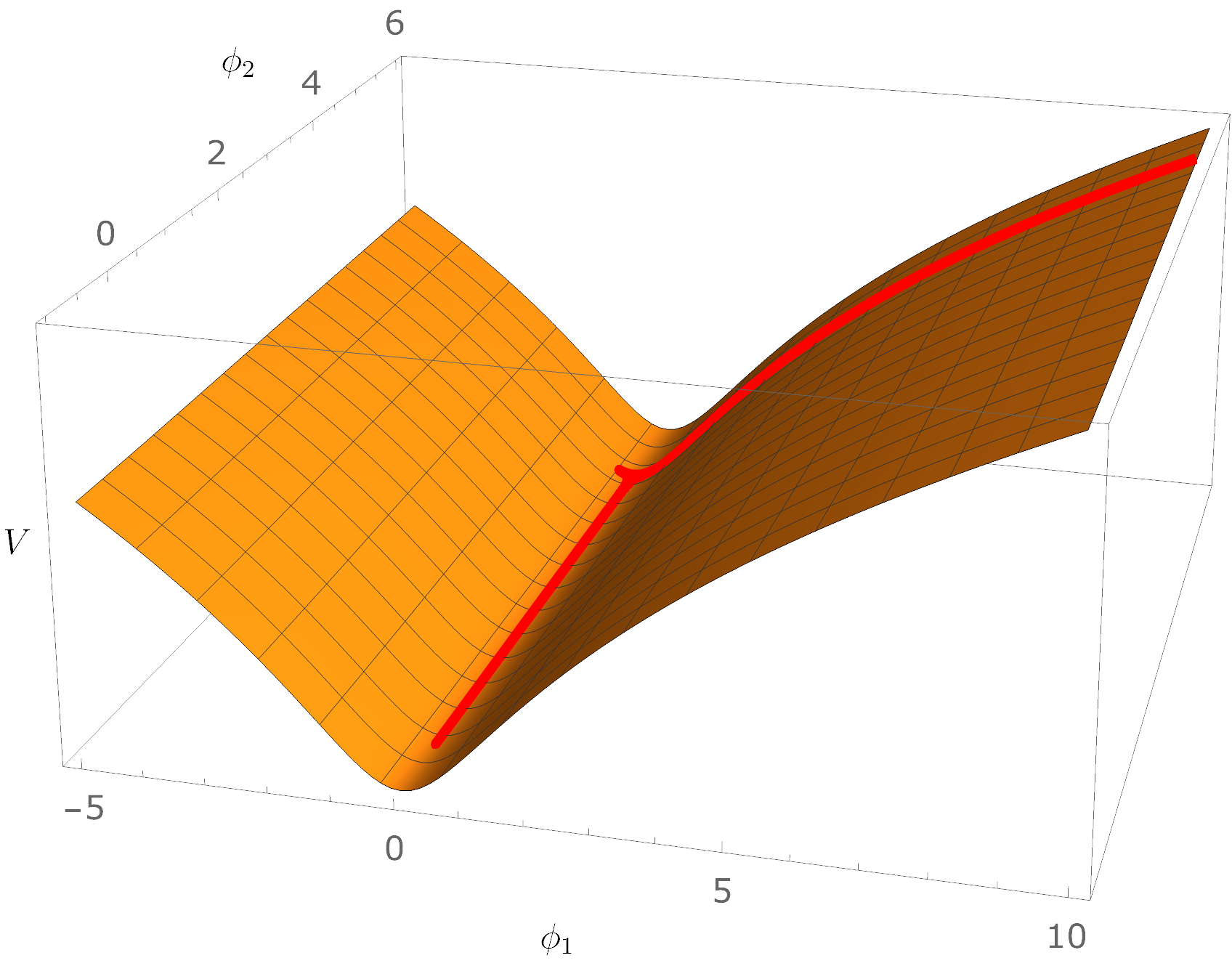}
    \caption{Two-field potential $V=V(\phi_1,\phi_2)$ for the model in eq. (2). 
    Here $M_2/M_1 = 0.1$, $p_1 = 2/5$, $p_2 = 1$, $\mu_1 = \mu_2 = {\cal O}(1)\,\Mpl$. 
    The red curve depicts a typical two-stage inflationary trajectory, where the field $\phi_1$ 
    slow-rolls down the slope first and then oscillates while decaying near the bottom of the valley. 
    Finally, $\phi_2$ starts to move along the valley.
    }
    \label{figpot}
\end{figure}

For the sake of simplicity, we will assume only 2 axions participating (see Fig.\ref{figpot})
\begin{equation}
V(\phi_1,\phi_2)= M_1^4 \left[\left(1+\frac{\phi_1^2}{\mu_1^2}\right)^{p_1/2}-1\right]
+ M_2^4 \left[\left(1+\frac{\phi_2^2}{\mu_2^2}\right)^{p_2/2}-1\right]\quad.
\end{equation}
The crucial point is that the $\sim 7$ observable efolds of the CMB will arise only 
from the part of inflation driven by the heavier axion. Hence, the CMB properties 
will be determined by the heavier axion slow-rolling only for $\Delta N_e < 60$ efolds. 
The other needed $60-\Delta N_e$ efolds are supplied after the break by the 2nd lighter axion. 
Since for large-field inflation models the CMB spectral index scales as $n_s = 1 - \mathcal{O}(1) / \Delta N_e$, 
we can have axion monodromy inflation with a good $n_s$ by doing double monodromy with two 
axions and placing the break at about 35 efolds before the end of inflation.

Incidentally, splitting up the full 60 efolds of slow-roll into several epochs driven 
by different axions has the effect of reducing the individual axion field 
displacement below the ${\cal O}(5 - 10)\,\Mpl$ needed for a single axion 
doing all the work. Given recent discussions in the literature about potential 
control issues for large trans-Planckian displacements of (pseudo)scalars due to the 
appearance of new states entering the original EFT, this may turn out to be a benefit.

For our purpose, the most interesting feature of this mechanism of rollercoaster 
axion monodromy is the possibility of one or several of the axions driving 
inflation to couple to a dark U(1) gauge field
\begin{equation}
{\cal L}_{\rm int}=\sqrt{-g} \frac{\phi_i}{4f_{\phi_i}} {F_i}_{\mu\nu}\tilde F_i^{\mu\nu}\quad.
\end{equation}
Here, we used again input from string constructions which seem to suggest that a 
given axion couples to at most a handful of different gauge fields, and often to 
just one (if at all -- often the U(1) gauge fields get Stueckelberg lifted and thus very massive).
The $f_{\phi_i}$ denote the axion decay constants.
For string axions, the decay constants arise from compactification of the extra dimensions and 
depend on inverse powers of the radii. As we assume these to be larger than the string scale, 
the decay constants are sub-Planckian.

The above coupling to gauge fields is interesting. Assume that the 
heavier of our 2 axions driving the 1st stage of inflation is coupled to a dark 
U(1) gauge field this way. Provided $f_{\phi_1}$ is small enough, the increasing 
speed of $\phi_1$ towards the end of its stage of inflation is known to generate a 
tachyonic-like instability for one of the 2 helicity eigenstates of the U(1) gauge field 
(see the appendix of \cite{Campbell:1992hc}). 
This leads to the generation of chiral gauge field quanta and quite drastic 
exponential production of this gauge field mode~\cite{Anber:2006xt,Anber:2009ua}. 
The maximum production level is bounded by the axion inflaton kinetic 
energy close to its slow-roll breaking exit point. 

Once these chiral gauge field quanta have been sourced, they in turn source a 
secondary peaked spectrum of stochastic gravitational waves~\cite{Barnaby:2011qe,pinky,Domcke:2017fix}. 
Their contribution to the energy density of the universe is given by
\begin{equation}
    \Om_{GW} \equiv \frac{\Om_{r,0}}{24} \Delta_T^2
    \simeq \frac{\Om_{r,0}}{12} \left(\frac{H}{\pi \Mpl}\right)^2
    \left( 1 + 4.3 \cdot 10^{-7} \frac{H^2}{\Mpl^2 \xi^6} e^{4 \pi \xi} \right) \quad .
\end{equation}
Here, we denote by $\Om_{r,0}=8.6\cdot 10^{-5}$ the present-day radiation 
density parameter. The quantity $\xi=\dot\phi_1 / (2H f_{\phi_1})$ controls the 
efficiency of the gauge field production and thus the strength of the overall effect. 
Strong gauge field production requires an axion decay small enough so that 
$\xi \gtrsim 3$ (see e.g.~\cite{DallAgata:2019yrr,Domcke:2020zez} for 
non-perturbative numerical treatments of the gauge field production and its 
backreaction on the inflationary dynamics).

\begin{figure}[ht]
    \centering
    \includegraphics[scale=0.8]{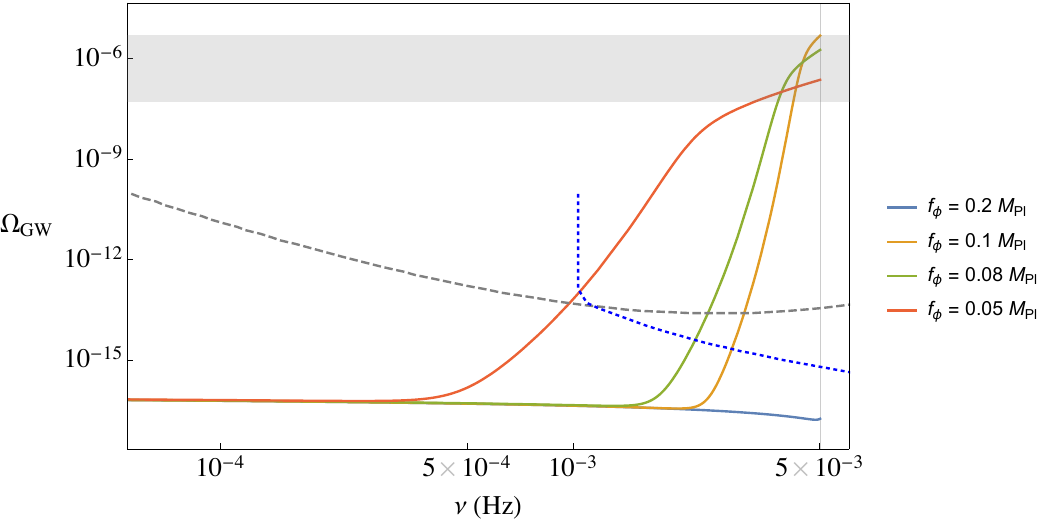}
    \caption{Abundance of gravitational waves as a function of frequency, setting $N_{\rm CMB}=\Delta N_e = 35$. 
    Dashed grey: the sensitivity of LISA. Dotted blue: the sensitivity of Big Bang Observer (BBO). 
    The example given here uses $M_1^4 = 2 \times 10^{-9} \,\Mpl^4$, $\mu_1 = \Mpl$, $p_1 = 0.2$.
    We scan $f_{\phi_1}$ over the values shown in the legend.}
    \label{fig:OmegaGWvaryf}
\end{figure}

This is the signal we are after. As the GW production traces the gauge field production, 
its peak frequency is determined by the number of efolds before the end of inflation 
when the first axion ends its inflationary stage and exits into the 2-3 efolds lasting break 
of oscillations providing matter domination. 
Using the relation between efold number and scale factor, and the fact that the comoving frequency $\nu = k/(2 \pi)$, we can express the frequency in terms of the number of efolds before the end of inflation:
\begin{equation}
    \label{eq:Ne_f}
    N = N_{\rm CMB} + \ln \frac{k_{\rm CMB}}{0.002 \, \Mpc^{-1}} - 44.9 - \ln \frac{\nu}{10^2 \, \Hz} \, \; ,
\end{equation}
where $N$, $N_{\rm CMB}$ are measured backwards from the moment of the end of 
inflation, when the signal peaks, and we fix $N_{\rm CMB}$ to be the time when CMB 
scales left the Hubble radius. Then $k_{\rm CMB} = 0.002 \, \Mpc^{-1}$ is the pivot scale.

We see that by choosing appropriate initial field values ${\phi_i}_0$ for our two axions, 
we can arrange the efold time $\Delta N_e$ of the break between the two stages of inflation. 
This in turn determines the frequency of the GW peak sourced by the gauge field production of the first 
heavier axion: if the first stage of inflation driven by $\phi_1$ lasts $\Delta N_e$ 
efolds before the break, then the observed CMB scales arise at $N_{\rm CMB}=\Delta N_e$ 
efolds before the end of the first stage.
An example of the expected spectrum for $\Delta N_e = 35$ 
and different $f_{\phi_1}$ is in Fig.\ref{fig:OmegaGWvaryf}. 
If we place the break at about $\Delta N_e = 50$ -- $51$ the 
resulting frequency range of the GW peak will fall into the $10 \, \kHz$ range of our levitating-sensor based detector.

\section{The axion GW signal}

We now look in more detail at the properties of the burst of primordial gravitational waves produced by double 
monodromy inflation coupled to a dark U(1) gauge field. This stochastic GW signal is peaked at some frequency 
$\nu_p$, which can vary (at production) from about few $\nHz$ to about $10 \, \MHz$, 
and with a width $\Delta \nu_p \simeq \frac{1}{10} \nu_p$.
This is a rough estimate, but it should be correct by order of magnitude.

In more detail, the signal can peak at different frequencies, depending on the number 
of efolds when the stage of inflation ends by decay into gauge fields.
Seeing such gravitational waves would show the existence of a dark $U(1)$ 
gauge sector, which is an amazing perspective.

\begin{figure}[ht]
    \centering
    \includegraphics[scale=0.5]{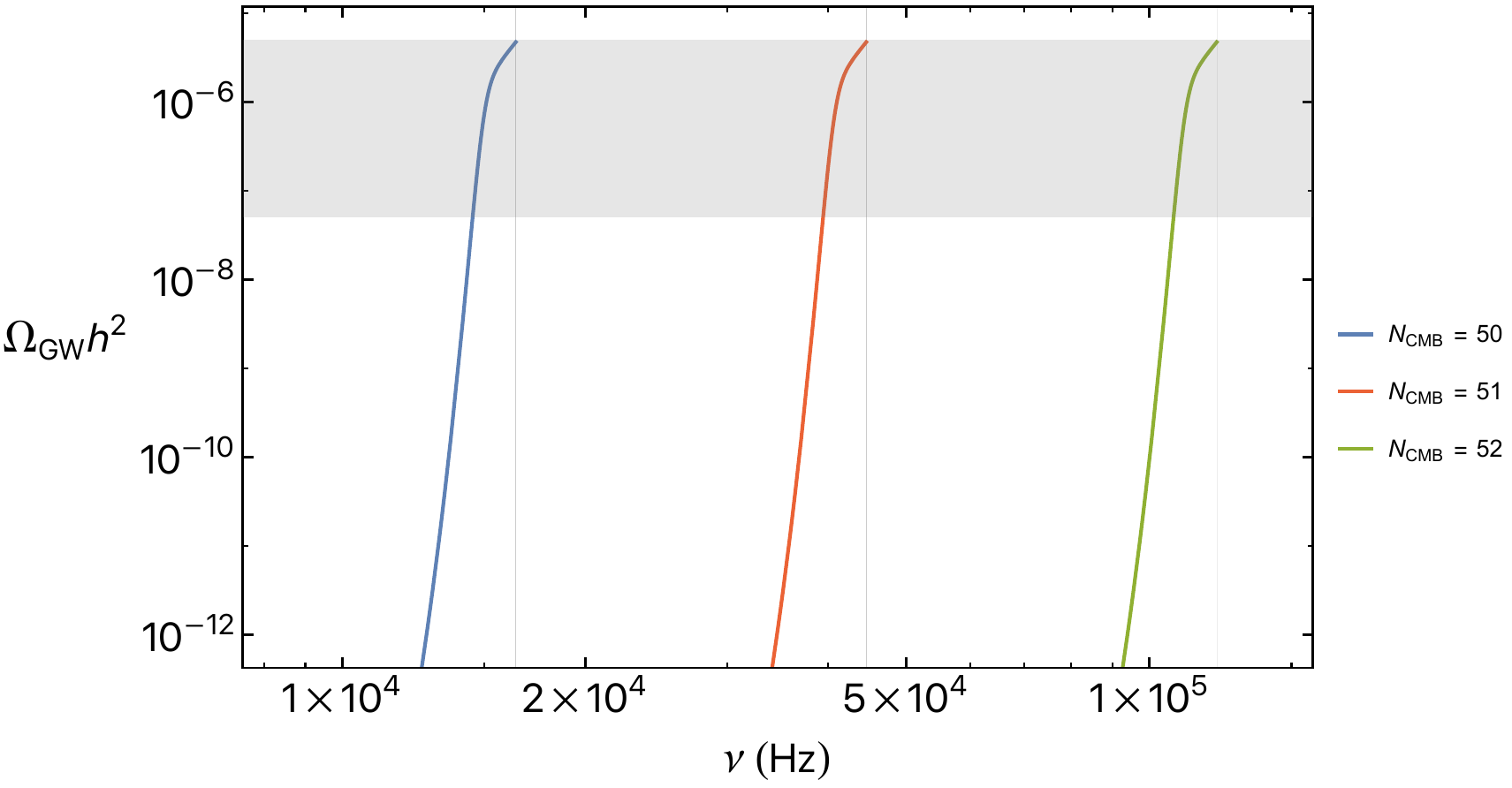}
    \caption{Abundance of gravitational waves as a function of frequency for different values of $N_{\rm CMB}$, 
    corresponding to the range of frequencies probed by the Levitated Sensor Detector.}
    \label{fig:OmegaGWvaryN}
\end{figure}

\begin{figure}[htb]
    \centering
    \includegraphics[scale=0.23]{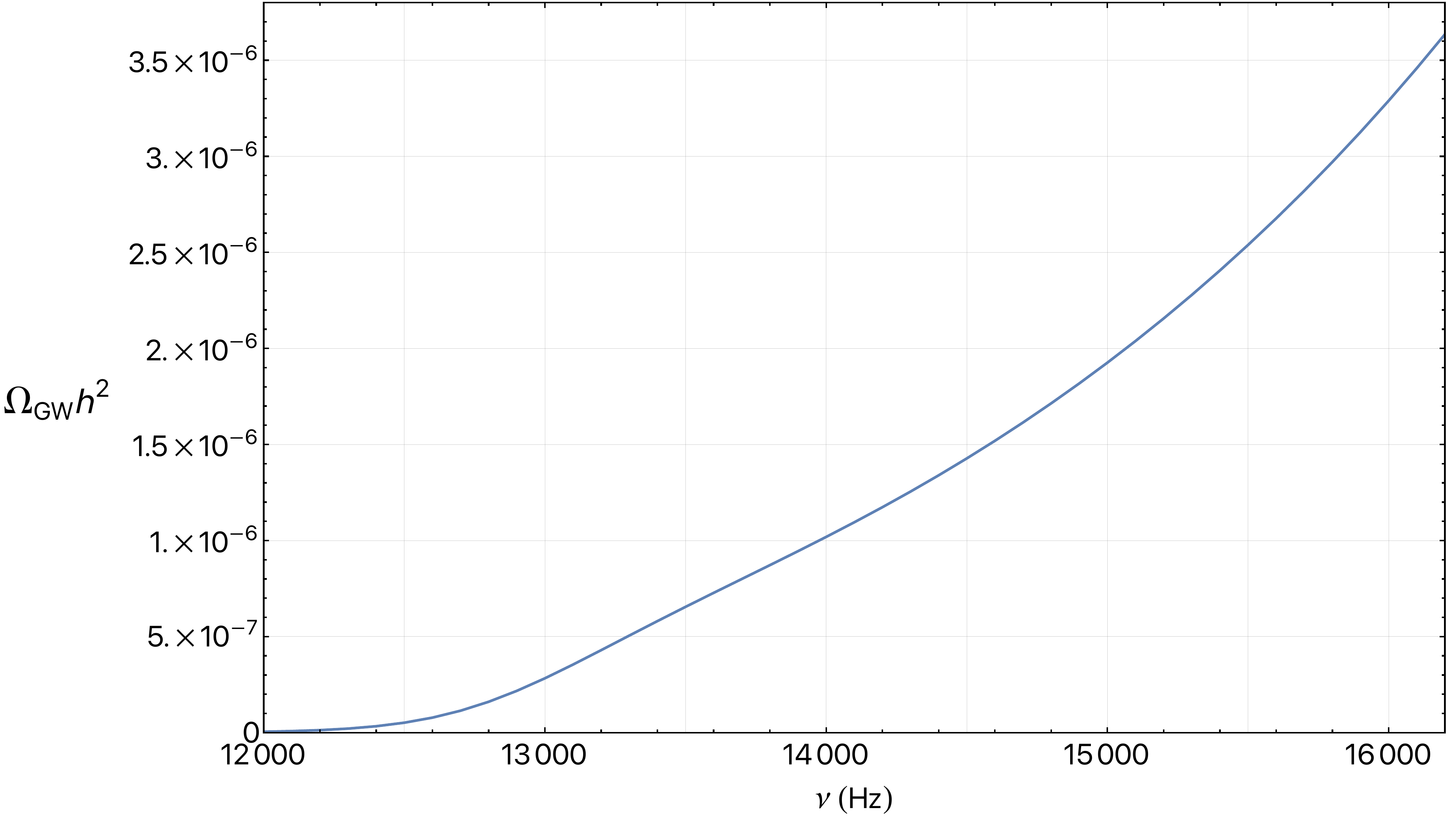}
      \caption{High-resolution plot of the rising edge of the gravitational wave peak as a function of frequency for 
      $N_{\rm CMB}=50$ corresponding to a $\sim 15 \, \kHz$ range signal.}
      \label{fig:deltaf}
\end{figure}

The range $10 \, \kHz$ -- $300 \, \kHz$ corresponds to a range of efolds $N_{\rm CMB} = 50$ -- $52$  
measured from the moment when the CMB scales left the Hubble radius, 
as seen in Fig.\ref{fig:OmegaGWvaryN}.
Here we have plotted a few different $N_{\rm CMB}$, with a fixed axion 
coupling, spanning the range of frequencies to which the Levitated Sensor Detector is sensitive. 

The grey area is an uncertainty, due to neglecting backreaction effects.
But this shows that we can reach the maximum value 
$\Omega_{GW}(\nu) h^2 \sim 5.7 \times 10^{-6} \Delta N_{\rm eff} \simeq 2 \times 10^{-6}$ 
allowed by bounds on $\Delta N_{\rm eff}$.
For the purpose of detection, we should determine the frequency width of the signal.
The $\Delta \nu/\nu$ is practically the same for all peaks, so we plot just 
one in Fig.\ref{fig:deltaf}, stopping at the $\Delta N_{\rm eff}$ bound.
The peaks are narrow, but not so much.
From Fig.\ref{fig:deltaf} we see that the width in frequency can be estimated as $\Delta \nu/\nu \simeq 0.1$.
With a full treatment, this may be larger, possibly up to $\Delta \nu/\nu \simeq 0.2$ 
which is what we estimate if we take the plot up to the full peak.

We can now look at the strain $h$, which is the quantity of interest in 
detection experiments. Here we make the assumption that our cosmological signal is 
stochastic, in the sense that the amplitudes $h(\nu)$ are random variables, of which we can predict correlators.
For estimates of the signals potentially observable in experiments, we make the following assumptions:
\begin{itemize}
    \item Stationarity (time-translation invariance). 
    This implies that $\avg{h^*(\nu) h(\nu')}$ is proportional to $\delta_D(\nu - \nu')$.
    \item Gaussianity. This means that the signal is characterized only by the 2-point function (assuming zero mean).
    \item Isotropic and not polarized background. This implies that we'll suppress angular 
    and polarization information.
\end{itemize} The signal can be modeled by defining the spectral density of the background
\begin{equation}
    \avg{h^*(\nu) h(\nu')} = \delta_D(\nu - \nu') \frac{1}{2} S_h(\nu) \, .
\end{equation}
What we predict is the energy density in GW, defined by
\begin{equation}
    \rho_{GW} = \Omega_{GW} \frac{3 c^2 H_0^2}{8 \pi G} = \frac{c^2}{32 \pi G} \avg{\dot{h}_{ij} \dot{h}^{ij}} \, .
\end{equation}
It is now convenient to write $\Omega_{GW}$ as an integral over log frequencies.
Given the definition of $S_h(\nu)$, we get
\begin{equation}
    \Omega_{GW} = \frac{1}{3 H_0^2} \int_{0}^{\infty} \rmd \ln \nu \, \nu \, (2 \pi \nu)^2 S_h(\nu) \, ,
\end{equation}
where a factor of $4$ comes from the sum over polarizations.
It is common to now define a dimensionless ``spectral density'' $\Omega_{GW}(\nu)$ by the equation
\begin{equation}
    \Omega_{GW} = \int_{0}^{\infty} \rmd \ln \nu \, \Omega_{GW}(\nu) \, .
\end{equation}

The strain spectral density $S_h$ is related to the dimensionless $\Omega_{GW}(\nu)$ (in frequency band $\nu$) by
\begin{equation}
    \Omega_{GW}(\nu) = \frac{4 \pi^2}{3 H_0^2} \nu^{3} S_h(\nu) \, .
\end{equation}
The $S_h$ has dimensions of inverse frequency, and it has a characteristic $1/\nu^3$ 
behavior for the scale invariant case in which $\Omega_{GW}(\nu)$ does not depend on frequency.
The quantity that commonly appears in the sensitivity plots is the strain $h(\nu) = \sqrt{S_h(\nu)}$, 
which has units of $\Hz^{-1/2}$, and is given by
\begin{equation}
    h(\nu) = \left( \frac{3 H_0^2}{4 \pi^2} \nu^{-3} \, \Omega_{GW}(\nu) \right)^{1/2 } \, .
\end{equation}
In sensitivity plots, as in our Fig.\ref{fig:hGWvsN}, it is common to plot a curve 
representing the bound on $\Delta N_{\rm eff}$ from BBN or CMB observations.
As mentioned in the introduction, it is important to note that such a curve represents 
the bound only in the case of a scale-invariant spectral density.
For our case, in which the $\Omega_{GW}(\nu)$ quantity is strongly peaked around 
a single frequency, the signal can cross the scale-invariant bound curve.
In fact, the measurements of $\Delta N_{\rm eff}$ translate in a bound on 
the full integral~\cite{Caprini:2018mtu}:
\begin{equation}
    \label{eq:Neffbound}
    \int_{\nu_{\rm BBN}}^{\infty} \rmd \ln \nu \, \Omega_{GW}(\nu) < 1.2 \times 10^{-5} \Delta N_{\rm eff} \, ,
\end{equation}
where $\nu_{\rm BBN} \simeq 10^{-12} \, \Hz$ is the frequency corresponding to the horizon 
size at BBN, and $\Delta N_{\rm eff} \lesssim 0.12$~\cite{ACT:2025tim,SPT-3G:2025bzu}. 
We have checked that our signal satisfies the bound.

\section{Estimating the sensitivity for the Levitated Sensor Detector}

The ground based laser interferometry gravitational observatories have attained a 
remarkable strain sensitivity to gravitational waves at frequencies below $10 \, \kHz$~\cite{aasi2015advanced}.
At higher frequencies, the sensitivity of such instruments is limited by photon shot noise.
Although advanced quantum sensing techniques such as squeezing can provide improvements 
for certain regimes of operation \cite{ligosqueeze}, in this higher frequency band, levitated 
optomechanical sensors provide a route towards substantially improved 
sensitivity \cite{Arvanitaki:2013,aggarwal_2022,aggarwal2025}.
In these instruments, a dielectric particle is suspended in an optical cavity 
and acts as a high-Q mechanical oscillator, with the gradient in the laser intensity 
along the optical standing wave providing the restoring force trapping the particle at 
an anti-node of the cavity field.
Due to the high mechanical $Q$ and relatively large displacement when driven on resonance, 
photon shot noise is not the dominant mechanism that limits the sensitivity.
Rather the sensitivity is limited by thermal noise in the motion of the particle, or in the 
extreme high vacuum limit by the radiation pressure shot noise associated with photon 
recoil heating of trap laser photons \cite{aggarwal_2022,winstone:2022}.
While the photon shot noise limited sensitivity of interferometer detectors 
such as LIGO becomes worse at higher frequency, the thermal noise limited 
sensitivity tends to improve, making the levitated sensor approach favorable 
with respect to an optical interferometer. The Levitated Sensor Detector currently 
under construction at Northwestern is a 1-meter prototype instrument intended to explore this technology.
The design sensitivity of the pilot instrument and future 10-meter or 100-meter long 
versions of the detector has been described in Ref. \cite{aggarwal_2022}.  
For the longest version considered previously, the strain sensitivity is of order 
$\sim 10^{-23}/\sqrt{\Hz}$, and would not be sufficient to detect the predicted 
signals within an observing time of $10^8$ seconds.

\begin{figure}[ht]
    \centering
    \includegraphics[scale=0.4]{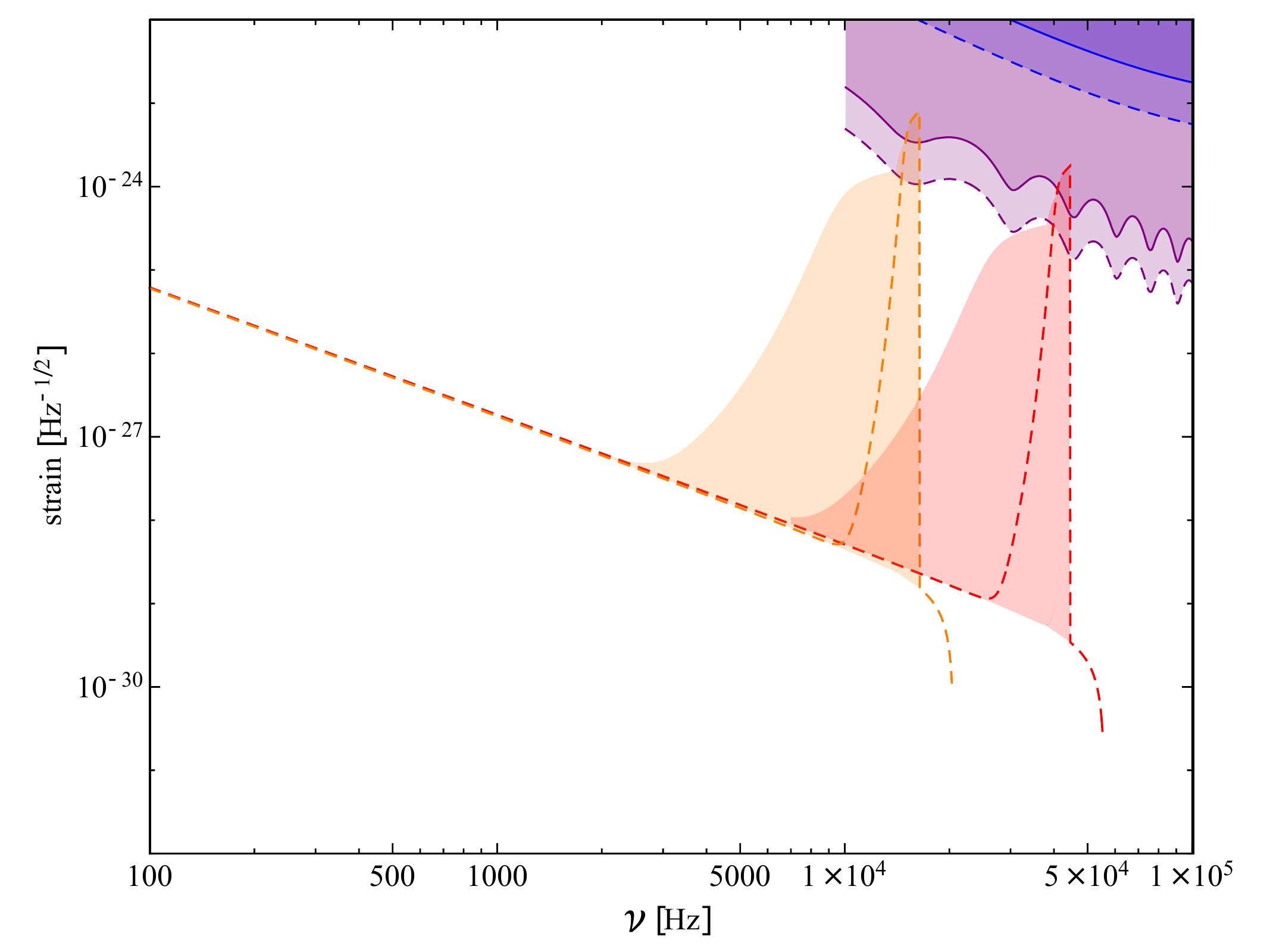}
    \caption{Estimated gravitational wave strain as a function of frequency, setting 
    different $N_{\rm CMB}=50, 51$ (orange, red). The dashed curve corresponds to 
    $f_\phi = 0.1 \Mpl$ and the shaded region represents an envelope with $f_\phi$ ranging 
    from $0.05$ to $0.15~\Mpl$.  Also shown are the sensitivity estimates of the 
    Levitated Sensor Detector for two different lengths $L = 100$ m (blue) and $L = 10$ km 
    (purple) for two different masses of the levitated particle, corresponding to the 
    system studied in Ref. \cite{aggarwal_2022} (solid) as well as three times larger mass (dashed).
    }
\label{fig:hGWvsN}
\end{figure}

However, it is possible to consider extending the length of such as setup to the 
10 kilometer scale of upcoming next generation observatories such as 
Einstein Telescope or Cosmic Explorer. One challenge of realizing this for a 
levitated optomechanics setup has to do with the need for a small mode waist.
The setup involves a Michelson interferometer configuration with Fabry-P\'erot arms 
of length $L$ as described in Ref. \cite{aggarwal_2022}, where a dielectric disk-shaped 
object is suspended at an anti-node of the standing wave inside each Fabry-P\'erot arm, 
where the optical restoring force results in a trapping frequency $\omega_0$. A second 
laser can be used to read out the position of the object as well as cool it along the cavity 
axes, as described for a similar setup in Ref. \cite{Arvanitaki:2013}. 
A passing GW with angular frequency $\omega_{GW}$ imparts a force on the 
trapped particle \cite{Arvanitaki:2013}, which is resonantly excited when 
$\omega_0=\omega_{GW}$. Unlike a resonant-bar detector, $\omega_0$ 
is widely tunable with laser intensity. 
The second cavity arm permits rejection of common mode noise, for example from 
technical laser noise or vibration.

The minimum detectable strain $h_{\rm{limit}}$ for a particle of mass $M$ 
with center-of-mass temperature $T_\mathrm{CM}$ is approximately \cite{Arvanitaki:2013}
\begin{equation}
h_{\mathrm{limit}}=\frac{4}{\omega_0^2L}\sqrt{\frac{k_BT_{\mathrm{CM}}\gamma_gb}{M}
\left[1+\frac{\gamma_{\mathrm{sc}}}{N_i\gamma_g}\right]}H\left(\omega_0\right), \label{eq:strain}
\end{equation}
where the cavity response function $H(\omega) = \sqrt{1+(2{\mathcal{F}}/\pi)^2\sin^2(\omega L /c)}$
for a cavity with finesse ${\mathcal{F}}$ and corresponding linewidth $\kappa = \pi c /L {\mathcal{F}}$. 
Here $N_i=k_BT_\mathrm{CM}/\hbar\omega_0$ is the mean initial phonon 
occupation number of the center-of-mass motion. $\gamma_g = \frac{32P}{\pi \bar{v} \rho t}$ 
is the gas damping rate at pressure $P$ with mean gas speed $\bar{v}$ for a disc of thickness 
$t$ and density $\rho$, and $b$ is the bandwidth.
The photon recoil heating rate \cite{Arvanitaki:2013,Jain:2016} 
$\gamma_{sc}=\frac{V_c\lambda\omega_0}{4L}\frac{1}{\int{dV(\epsilon-1)}}\frac{1}{\mathcal{F}_{\rm{disc}}}
$ is inversely proportional to the disc-limited finesse $\mathcal{F}_{{\rm{disc}}}$, 
i.e. ($2\pi$) divided by the fraction of photons scattered by the disc outside the cavity mode. 
The integral is performed over the extent of the suspended particle. Here $V_c$ is the 
cavity mode volume \cite{Arvanitaki:2013}, $\epsilon$ is the particle dielectric constant, 
and $\lambda$ is the trapping laser wavelength. 

For the application of GW detection, using high-mass, 
high-trap-frequency, disc- or plate-like microparticles is ideal for 
achieving maximal sensitivity with this technique. Recent experimental work 
has demonstrated optical trapping of high-aspect-ratio hexagonal prisms with a 
20 micron-scale radius \cite{winstone:2022}. The prisms are trapped in vacuum using 
an optical standing wave, with the normal vector to their face oriented along the beam 
propagation direction, yielding much higher trapping frequencies than those typically achieved 
with microspheres of similar mass. 
This plate-like geometry is planned to be used in high frequency gravitational 
wave searches in the Levitated Sensor Detector, currently under construction \cite{aggarwal_2022}.
We consider a setup with parameters as described in Tables~\ref{tab:table1} and~\ref{tab:table2}, 
corresponding to different cavity lengths, finesse, and masses for the levitated dielectric stack. 
Relative to the setup described in prior work \cite{aggarwal_2022}, the longer (10 km) configuration 
would need an additional lens to focus the beam near the input mirror of each arm cavity, 
to provide sufficient transverse restoring force to support the particle against the earth's gravity. 
We assume that the modest cavity finesse shown in Tables~\ref{tab:table1} and~\ref{tab:table2} 
can be achieved in the presence of such a lens and that with sufficient vibration isolation and 
use of low-loss materials, the contribution of the lens to the noise budget of the system can be 
made comparable to the Fabry-P\'erot end mirrors.

\begin{table}[!t]
\begin{center}
  \begin{tabular}{@{}cccccc@{}}
  \hline
  \hline
   Parameter & Units & \hspace{5pt} $L=100$ m \hspace{5pt}  & \hspace{5pt} $L=100$ m 
   \hspace{5pt} & \hspace{5pt} $L=10$ km \hspace{5pt} & \hspace{5pt} $L=10$ km \hspace{5pt} \\
  \hline
$\omega_0/2\pi$ & kHz & $10$ & $100$ & $10$ & $100$\\
$I_0$ & W/$m^2$ & $2.2 \times 10^8$ & $2.2 \times 10^{10} $ & $2.2 \times 10^8$ & $2.2 \times 10^{10} $\\
${\mathcal{F}}$ & -- & $10$ & $10$ & $3$ & $3$\\
$N_i \gamma_g $ & Hz & $0.54$ & $0.054$ & $0.54$ & $0.054$\\
$ \gamma_{sc} $ & Hz & $0.004$ & $0.041$ & $0.004$ & $0.041$\\
$h_{\rm{min}}$ & 1$/\sqrt{\Hz}$ & $7.9 \times 10^{-22}$ & 
$1.7 \times 10^{-23}$ & $1.5 \times 10^{-23}$ & $2.0 \times 10^{-25}$ \\
  \hline
  \hline
  \end{tabular}
\caption{Experimental parameters for trapping of a $75$ $\mu$m 
radius stack with $14.58$ $\mu$m thick SiO$_2$ spacer (corresponding to $j=28$) 
and quarter-wave $110$ nm thick Si endcaps in a cavity of length $L$ at $P=10^{-11}$ Torr 
and room temperature. $I_0$ is the peak laser intensity striking the disc and 
$h_{\rm{min}}=h_{\rm{limit}}/\sqrt{b}$ is the strain sensitivity where $b$ is 
the measurement bandwidth.}
\label{tab:table1}
\end{center}
\end{table}

\begin{table}[!t]
\begin{center}
  \begin{tabular}{@{}cccccc@{}}
  \hline
  \hline
   Parameter & Units & \hspace{5pt} $L=100$ m \hspace{5pt}  & 
   \hspace{5pt} $L=100$ m \hspace{5pt} & \hspace{5pt} $L=10$ km 
   \hspace{5pt} & \hspace{5pt} $L=10$ km \hspace{5pt} \\
  \hline
$\omega_0/2\pi$ & kHz & $10$ & $100$ & $10$ & $100$\\
$I_0$ & W/$m^2$ & $2.2 \times 10^8$ & $2.2 \times 10^{10} $ & $2.2 \times 10^8$ & $2.2 \times 10^{10} $\\
${\mathcal{F}}$ & -- & $10$ & $10$ & $3$ & $3$\\
$N_i \gamma_g $ & Hz & $0.17$ & $0.017$ & $0.17$ & $0.017$\\
$ \gamma_{sc} $ & Hz & $0.0013$ & $0.013$ & $0.0013$ & $0.013$\\
$h_{\rm{min}}$ & 1$/\sqrt{\Hz}$ & $2.5 \times 10^{-22}$ & 
$5.5 \times 10^{-24}$ & $4.8 \times 10^{-24}$ & $6.4 \times 10^{-26}$ \\
  \hline
  \hline
  \end{tabular}
\caption{Experimental parameters for trapping of a $75$ $\mu$m 
radius stack with $45.82$ $\mu$m thick SiO$_2$ spacer (corresponding to $j=88$) 
and quarter-wave $110$ nm thick Si endcaps in a cavity of length $L$ at $P=10^{-11}$ Torr 
and room temperature. $I_0$ is the peak laser intensity striking the disc and 
$h_{\rm{min}}=h_{\rm{limit}}/\sqrt{b}$ is the strain sensitivity where $b$ is the measurement bandwidth. }
\label{tab:table2}
\end{center}
\end{table}

In Fig.\ref{fig:hGWvsN}, we illustrate the sensitivity for an integration 
time of $10^8$ seconds for two particular cases of a signal corresponding $N_{\rm CMB}=50, 51$.  
Here we illustrate one particular value of $f_\phi = 0.1 \Mpl$ as a dotted curve and show an 
envelope obtained by varying the value from $f_\phi=0.05 \Mpl$ to $f_\phi=0.15 \Mpl$. 
This envelope provides us with a rough estimate of the `theory error' encapsulating a UV theory 
motivated range of the otherwise unknown value of the axion decay constant $f_\phi$.
The cavity response function can be approximated as 
$H\left(\omega\right) \approx \sqrt{1+4\omega^2/\kappa^2}$ for a sufficiently short cavity, e.g. 
valid for the blue solid and dashed curves shown in Fig.\ref{fig:hGWvsN} for the 100 m long cavity. 
For the longer 10 km arm cavity we consider, the oscillations of the sine function 
are visible in the estimated sensitivity curve.

\emph{Considerations for multiple detectors.}
For a single detector, the presence of a stochastic GW background signal 
would manifest essentially as excess noise in the relevant frequency band 
for example as shown in Fig.\ref{fig:hGWvsN}. 
Defining a factor $F$ that takes into account the angular efficiency of the detector,
the signal-to-noise ratio, for each frequency $\nu$ is
\begin{equation}
    \left(\frac{S}{N}\right)^2(\nu) = F \frac{S_h(\nu)}{S_n(\nu)} \ 
\end{equation} where $S_n(\nu)$ is the noise spectral density.
Supposing the noise spectrum is essentially flat over the bandwidth of the signal, 
we can consider integrating over a frequency window of $\Delta \nu \lesssim \Delta \nu_p$, 
corresponding to an observational time $T \gtrsim \frac{1}{\Delta \nu_p}$. Using the frequencies 
in the plot, for the smallest $\nu_p \simeq 10 \, \kHz$, we require 
$T \gtrsim \frac{1}{\Delta \nu_p} \simeq 10^{-3} \, \s$. Then for averaging times up 
to $10^{-3}$ the signal is essentially phase coherent whereas the noise is incoherent 
e.g. for a thermal noise background, so that the signal to noise ratio improves, scaling as 
$T^{1/2}$. For longer averaging times, e.g. such as the $10^8$ s considered in Fig.\ref{fig:hGWvsN}, 
the phase of the signal varies as the time duration exceeds the coherence time, 
and the scaling of the improvement with averaging time is reduced to $T^{1/4}$, as is 
typically the case for detection of other stochastic strain-related signals such as that 
resulting from dilatonic ultralight dark matter \cite{derevianko2018}. 

When we have two detectors, we can try and correlate the outputs to 
separate the (correlated) signal from the noise,
in particular for two detectors that are separated by a distance $\Delta x$ 
small compared to the gravitational wave wavelength, i.e. $\Delta x \lesssim \frac{c}{2 \pi \nu}$.
To estimate this range, take now the largest frequencies of the plot, 
$\nu_{\rm max} \simeq 100 \, \kHz$, this requires
\begin{equation}
    \Delta x \lesssim \frac{3 \times 10^8}{2 \pi \times 10^{5}} \, \m \simeq 400 \, \m \, ,
\end{equation}
which is reasonably achievable for $100 \, \m$ long detector arms. In this 
regime we expect a factor of $\sqrt{2}$ improvement in the sensitivity due to the 
correlated signal, which scales as $\sqrt{N}$ for $N$ instruments.
At $10 \, \kHz$ this distance increases to 4 km. For the 10 km baseline taken for the 
largest scale instrument we consider, this length is larger than the gravitational wave 
wavelengths being considered. 

\section{Summary}

In this paper we have considered a class of inflationary scenarios in which accelerated 
expansion proceeds in multiple stages, driven by several inflaton sectors separated by 
mass hierarchies. Rather than realizing the required $\sim 60$ efolds of inflation in a single 
continuous phase, inflation is intermittently interrupted as the dominant inflaton sector 
changes. This setup naturally arises in frameworks with multiple axions or braneworld 
constructions and avoids some of the structural challenges faced by single-field models, 
while also evading the standard inflationary no-hair arguments that typically erase 
observable imprints of pre-inflationary or transient dynamics.

A key consequence of such interruptions is the generation of distinctive gravitational 
wave signals. In models where axion-like inflatons couple to dark $U(1)$ gauge fields, 
the end of an inflationary stage can trigger exponentially enhanced production of gauge 
modes, which rapidly source chiral gravitational waves. Unlike the nearly scale-invariant 
stochastic background expected from vacuum fluctuations, these signals are sharply 
localized in frequency. Earlier work has shown that interruptions occurring $\mathcal{O}(10)$ 
efolds before the end of inflation can produce strong gravitational wave backgrounds at 
subhorizon scales, potentially accessible to current and future experiments.

Focusing on interruptions occurring only a few efolds before the end of the final inflationary stage, 
we find that the resulting gravitational waves today populate the high-frequency band 
$\nu \sim 10^{4} - 5\times10^{5} \, \Hz$, corresponding to wavelengths of order $100\,\mathrm{m}$ 
to tens of kilometers. These frequencies lie in a regime where contamination from late-universe 
astrophysical and cosmological sources is minimal. Remarkably, the predicted strain can approach 
the sensitivity thresholds of proposed terrestrial detectors such as the Einstein Telescope, Cosmic 
Explorer, and Levitated Sensor Detector experiments. Because the signal is extremely narrow in 
frequency, its total energy density easily satisfies existing bounds from BBN and other 
gravitational wave constraints, even when the peak amplitude greatly exceeds that of competing sources.

These results identify a qualitatively new observational window on inflationary physics. 
Multi-stage inflation with brief interruptions can leave pronounced, localized gravitational 
wave signatures that are both theoretically motivated and experimentally accessible, offering a 
promising avenue for probing the microphysics of inflation well beyond the reach of 
conventional cosmological observables.

\vskip.75cm

{\bf Acknowledgments}: 
N.K. and A.W. would like to thank ``Schweinske" Bahrenfeld for providing a hospitable  
atmosphere and a free WIFI connection. 
AG acknowledges support from NSF grants PHY-2409472 and PHY-2111544, DARPA, 
the John Templeton Foundation, the W.M. Keck Foundation, the Simons Foundation, 
the Alfred P. Sloan Foundation under Grant No.\ G-2023-21130, the Gordon and Betty Moore Foundation 
Grant GBMF12328, DOI 10.37807/GBMF12328.  NK is supported in part by the DOE Grant DE-SC0009999. 
AW is partially supported by the Deutsche Forschungsgemeinschaft under Germany's 
Excellence Strategy - EXC 2121 ``Quantum Universe" - 390833306, by the 
Deutsche Forschungsgemeinschaft through a German-Israeli Project Cooperation (DIP) grant 
``Holography and the Swampland", and by the Deutsche Forschungsgemeinschaft 
through the Collaborative Research Center SFB 1624 ``Higher Structures, 
Moduli Spaces, and Integrability". 

\bibliographystyle{utphys}
\bibliography{references}

\end{document}